\documentclass[a4paper]{spie}  

\usepackage{subfigure}
\usepackage{amsmath}
\usepackage{amssymb}
\usepackage{amsfonts}
\usepackage{bm}
\usepackage{color}
\usepackage[colorlinks=true,citecolor=blue,urlcolor=blue,linkcolor=blue,hyperfigures=true]{hyperref}

\newif\ifpdf
\ifx\pdfoutput\undefined
   \pdffalse
\else
   \pdfoutput=1
   \pdftrue
\fi
\ifpdf
   \usepackage{graphicx}
   \usepackage{epstopdf}
\else
   \usepackage{graphicx}
\fi

\graphicspath{{./Figures/}}
\def\Eq{Eq.~}

\def\Fig{Fig.~}

\def\Ref{Ref.~}
\def\Refs{Refs.~}
\def\Sec{Sec.~}

\def\Tab{Table~}
\def\be{\begin{equation}}
\def\ee{\end{equation}}
\def\bea{\begin{eqnarray}}
\def\eea{\end{eqnarray}}

\def\ie{\textit{i.e.}~}
\def\eg{\textit{e.g.}~}

\def\dd{\mathrm{d}}

\title{Development of compact cold-atom sensors\\for inertial navigation}

\author[a,$*$]{B. Battelier}
\author[a,b]{B. Barrett}
\author[a]{L. Fouch\'{e}}
\author[a]{L. Chichet}
\author[a]{L. Antoni-Micollier}
\author[b]{H. Porte}
\author[b]{F. Napolitano}
\author[c,$\dagger$]{J. Lautier}
\author[c]{A. Landragin}
\author[a]{P. Bouyer}
\affil[a]{LP2N, IOGS, CNRS and Universit\'e de Bordeaux, rue Fran\c{c}ois Mitterrand, 33400 Talence, France}
\affil[b]{iXBlue, 34 rue de la Croix de Fer, 78100 Saint-Germain-en-Laye, France}
\affil[c]{LNE-SYRTE, Observatoire de Paris, PSL Research University, Sorbonne Universit\'{e}s, UPMC Univ. Paris 06, 61 avenue de l'Observatoire, 75014 Paris, France}

\authorinfo{$^*$Corresponding author's email: baptiste.battelier@institutoptique.fr\\
$^\dagger$Present address: muQuanS, Institut d'Optique d'Aquitaine, rue Fran\c{c}ois Mitterrand, 33400 Talence, France}

\begin{document}

\maketitle

\begin{abstract}
Inertial sensors based on cold atom interferometry exhibit many interesting features for applications related to inertial navigation, particularly in terms of sensitivity and long-term stability. However, at present the typical atom interferometer is still very much an experiment---consisting of a bulky, static apparatus with a limited dynamic range and high sensitivity to environmental effects. To be compliant with mobile applications further development is needed. In this work, we present a compact and mobile experiment, which we recently used to achieve the first inertial measurements with an atomic accelerometer onboard an aircraft. By integrating classical inertial sensors into our apparatus, we are able to operate the atomic sensor well beyond its standard operating range, corresponding to half of an interference fringe. We report atom-based acceleration measurements along both the horizontal and vertical axes of the aircraft with one-shot sensitivities of $2.3 \times 10^{-4}\,g$ over a range of $\sim 0.1\,g$. The same technology can be used to develop cold-atom gyroscopes, which could surpass the best optical gyroscopes in terms of long-term sensitivity. Our apparatus was also designed to study multi-axis atom interferometry with the goal of realizing a full inertial measurement unit comprised of the three axes of acceleration and rotation. Finally, we present a compact and tunable laser system, which constitutes an essential part of any cold-atom-based sensor. The architecture of the laser is based on phase modulating a single fiber-optic laser diode, and can be tuned over a range of 1 GHz in less than 200 $\mu$s.
\end{abstract}


\section{INTRODUCTION}
\label{sec:Intro}

Cold atoms have been demonstrated to be very effective tools for high-precision tests of fundamental physics, as well as for applications which require very accurate measurements of frequencies \cite{Markowitz-PRL-1958, Parker-Metrologia-2010, Marion-PRL-2003, Laurent-EPJD-1998} or inertial forces such as acceleration, gravity, or rotation \cite{Peters-Nature-1999, Bertoldi-EPJD-2006, Snadden-PRL-1998, Gustavson-PRL-1997, Canuel-PRL-2006, Altin-NJP-2013, Hardman-arXiv-2016}. Until recently, the complexity of these experiments made it very difficult to imagine commercial systems used by non-specialists. During the past few years, many efforts to develop the technology have been carried out to simplify, integrate and reduce the size of cold-atom setups. Nowadays, atom interferometers and atomic clocks are starting to be available as black-box devices for industrial applications on ground \cite{AOSense-Website, muQuans-Website} and position themselves as next-generation instruments for precision measurements in Space \cite{Laurent-EPJD-1998, Schuldt-ExpAst-2015}. The availability of compact inertial sensors based on atom interferometry could potentially lead to a breakthrough in quantum technology for future generations of accelerometers and gyroscopes with ultra-high accuracy. Many applications exist for this type of instrument, such as geophysical research, underground prospecting, civil engineering, inertial navigation or gravitational physics \cite{Barrett-VarennaProceedings-2014}, and the demand for higher performance is only expected to increase. However, the use of cold-atom sensors for positioning and guidance still faces many scientific and technological challenges, such as compliance with onboard applications, compactness, robustness against environmental effects, measurement continuity, and the potential for interference with other instrumentation. In this article, we present recent developments on these fronts from various projects underway at the Laboratoire Photonique, Num\'{e}rique et Nanoscience (LP2N) in Bordeaux, France.


The joint laboratory iXAtom brings together the knowledge of the French company iXBlue---experts in optical gyroscopes, photonics and inertial navigation---and personnel from LP2N specialized in atom interferometry. The aim of this collaboration is to make technological advances using cold atoms to develop the next generation of inertial sensors for industrial, military and Space applications, with anticipated improvements in their performance. In the near future, we plan to develop a compact gyroscope and a three-axis accelerometer based on new techniques in atom interferometry. The ultimate goal of iXAtom is to build a new autonomous device which can compete with technologies based on global positioning systems (GPS) without the drawback of external communication for recalibration.

This new project is based on previous work done at LP2N. First, the ICE (Interf\'{e}rom\'{e}trie atomique \`{a} sources Coh\'{e}rentes pour l'Espace) project is an ongoing experiment designed to make a precise test of the Universality of Free Fall using the two atomic species $^{87}$Rb and $^{39}$K \cite{Barrett-VarennaProceedings-2014, Barrett-NJP-2015}. It is also designed as a prototype of a mobile atomic accelerometer, which recently demonstrated the first onboard measurements of inertial effects in an aircraft \cite{Stern-EPJD-2009, Geiger-NatComm-2011}. An interesting feature of ICE is that the experiment can be configured to be sensitive to accelerations along the horizontal or vertical axes. In \Sec \ref{sec:OnboardICE}, we review recent acceleration measurements during parabolic flights onboard the Novespace Zero-G aircraft. In the near future, we also plan to implement multi-axis inertial measurements using the same setup, which we discuss in \Sec \ref{sec:3DCAI}. To reach the goals of iXAtom, building a compact and simple laser system is a key feature. Pioneering work has already been done in this direction on the MiniAtom project whose aim was to build a compact prototype to measure gravity with cold atoms. In the frame of this study, a fast, tunable and compact system based on a single fibered laser diode operating in the telecom frequency band has been developed. We describe the architecture of MiniAtom laser in detail in \Sec \ref{sec:FastTunableLaser}.

\section{Principles of inertial navigation}
\label{sec:InertialPrinciples}

The goal of inertial navigation is to track the position of an object over an arbitrary spatial trajectory using only local measurements of the object's acceleration and rotation vectors. In practice, this task is quite challenging due to errors in position and orientation that diverge with time. Nowadays, with the broad availability of GPS devices, navigation is typically challenging only where GPS signals cannot penetrate---in sub-surface or sub-marine locations, for instance. Furthermore, even the most precise GPS devices exhibit position resolutions of a few meters, and cannot give any information regarding an object's orientation. For these reasons, independent and autonomous methods of positioning, such as inertial navigation, are of great interest to many industries.

The principle of inertial navigation is as follows. A complete inertial measurement unit (IMU), composed of three mutually-orthogonal accelerometers and gyroscopes, is attached to a moving object. A computer continuously acquires the vector acceleration and rotation output by the IMU, and the position and orientation of the object is computed. Nowadays, virtually all numerical procedures used for inertial positioning are some variation of the ``strapdown'' navigation algorithm \cite{Adams-JNav-1956, Titterton-Book-2004, Woodman-Report-2007, Lefevre-Book-2014}. This procedure involves first integrating the rotation rates measured in the frame of the moving body to obtain an orientation in the inertial frame at time $t_0$. The global orientation is used to project the accelerations measured in the body frame onto the inertial frame. After subtracting the acceleration due to gravity, the global accelerations are then integrated once to obtain the velocity, and again to obtain the position of the object at some later time $t_0 + \Delta t$. Repeating this procedure for each measurement output by the IMU yields the vector position and orientation as a function of time. In conjunction with the knowledge of the initial position and orientation, all the information concerning the trajectory of the object is known.

\subsection{Performances of standard inertial navigation}
\label{sec:ClassicalNavigation}

The accelerometers contained within an IMU are typically mechanical in nature, and behave as a damped mass on a spring. When accelerated, the mass is displaced to the point where the spring compensates the associated force, and the displacement is measured to obtain the acceleration. In commercial devices, piezo-electric, piezo-resistive and capacitive components are commonly used to convert the mechanical motion into an electrical signal. Nowadays, very compact accelerometers are achieved with micro-electro-mechanical systems (MEMS). Accelerometers with ultra-high performances \cite{Josselin-SensActA-1999, Haeussermann-PIGA-2001} have also been developed for specific applications, but at a very expensive cost. The main limitation for standard accelerometers is long-term bias stability, which causes the error in position to rapidly diverge.

As for measuring the rotation rate, two technologies dominate the market for navigation-grade gyroscopes: the ring-laser gyroscope (RLG) and the fiber-optic gyroscope (FOG), which are both based on the recirculating Sagnac effect. While both RLGs and FOGs have demonstrated similar performance---reaching stabilities better than $10^{-3}$ deg/h ($5 \times 10^{-9}$ rad/s)---RLGs, in their current form, have reached their limitation due to the Fresnel-Fizeau drag effect \cite{Lefevre-Book-2014}. On the other hand, FOGs have continuously benefitted from advances in the optical-fiber communications industry since the revolutionary development of the erbium-doped fiber amplifier (EDFA) in the 1990s. But in a similar manner to accelerometers, they are limited at long timescales by bias drift, and have not yet reached their ultimate performance goal of $10^{-5}$ deg/h.

\begin{table}[!tb]
  \centering
  \small
  \begin{tabular}{|c|cc|cc|}
    \hline
      Sensor  & Accelerometer bias & Position      & Gyroscope bias    & Position  \\
      grade   & stability (m$g$)   & accuracy (km) & stability (deg/h) & accuracy (km) \\
    \hline
      Consumer     & $>50$   & $>320$  & $>100$     & $>11000$ \\
      Industrial   & $10$    & $64$    & $10 - 100$ & $1100 - 11000$ \\
      Tactical     & $1$     & $6.4$   & $1 - 10$   & $110 - 1100$ \\
      Intermediate & $0.1$   & $0.64$  & $0.01 - 1$ & $1.1 - 110$ \\
      Navigation   & $0.05$  & $0.32$  & $<0.01$    & $<1.1$ \\
      Strategic    & $0.005$ & $0.032$ & $<0.001$   & $<0.11$ \\
    \hline
  \end{tabular}
  \caption{General classification of accelerometer and gyroscope performance (adapted from \Ref \citenum{Lefevre-Book-2014}). Here, the bias stabilities are characterized over 1 h of integration (1 deg/h $= 4.8 \times 10^{-6}$ rad/s). The position accuracy associated with the accelerometer bias is computed based on the amplitude of the 84-minute Schuler oscillations given by \Eq \eqref{eq:Schuler}. For the gyroscopes, the position accuracy is estimated using $\delta\Omega R_{\oplus} t$ for a period of $t = 1$ h, where $R_{\oplus}$ is the radius of the Earth, and $\delta \Omega$ is the bias stability.}
  \label{tab:SensorGrades}
\end{table}

Table \ref{tab:SensorGrades} shows a general classification of inertial sensors in terms of their bias stability. Even the most sensitive class of commercial sensors produces position uncertainties of hundreds of meters after only 1 h of navigation time---much larger than the 10 m--scale resolution of commercial GPS units. It is therefore common practice to enhance the level of accuracy of inertial navigation systems by integrating measurements from external ``non-inertial'' sensors, such as odometers, barometers, GPS, or acoustic positioning systems. The data fusion process is most commonly performed using modern signal processing techniques---Kalman filtering being one of the most popular methods \cite{Kalman-JBE-1960}.

\subsection{Using cold atom sensors for inertial navigation}
\label{sec:ColdAtomNavigation}

The principle of a cold-atom interferometer (CAI) is to measure the position of an atom relative to the phase ruler defined by a retro-reflected laser beam. A typical interferometer sequence is composed of three light pulses---each of which imprints the phase of the laser $\Phi(t_i) = \bm{k}_{\rm eff} \cdot \bm{r}(t_i) - \omega_{\rm eff} t_i + \phi_i$ at a given time $t_i$ onto the atomic wavefunction. Here, $\bm{k}_{\rm eff}$ and $\omega_{\rm eff}$ are the effective wavevector ($\bm{k}_{\rm eff} = \bm{k}_1 + \bm{k}_2$) and frequency ($\omega_{\rm eff} = \omega_1 - \omega_2$) of the Raman light, $\phi_i$ is an arbitrary laser phase, and $\bm{r}(t)$ represents the trajectory of the atom (defined with respect to the reference frame of the mirror used to reflect the light). For a Mach-Zehnder geometry, each successive pulse is separated by an interrogation time $T$, and the phase difference between interfering atoms is $\Delta\Phi = \Phi(0) - 2\Phi(T) + \Phi(2T)$. Thus, for an atom undergoing a simple accelerating trajectory $\bm{r}(t) = \bm{r}_0 + \bm{v}_0 t + \frac{1}{2} \bm{a} t^2$, the total phase difference can be shown to be
\begin{equation}
  \label{eq.PhiAccelero}
  \Delta\Phi = \bm{k}_{\rm eff} \cdot \bm{a} T^2 + \Delta\phi,
\end{equation}
where $\Delta\phi = \phi_1 - 2\phi_2 + \phi_3$ is the total laser phase, which is typically used as a control parameter to scan the interference fringe. Since a CAI is sensitive to accelerations only along the axis of the laser, it can be configured such that phase given by \Eq \eqref{eq.PhiAccelero} is dominated by the acceleration due to gravity ($\bm{a} = \bm{g}$) in the case of a gravimeter, or the Coriolis acceleration ($\bm{a} = -2\bm{\Omega}\times\bm{v}$) in the case of a gyroscope. Intuitively, a CAI outputs a phase that is proportional to the acceleration of the atoms relative to the reference mirror attached to the device. Thus, for a CAI in a moving vehicle, what is measured is the difference between the perfect free-fall trajectory of the atoms and the actual trajectory of the vehicle.

Cold-atom accelerometers (CAAs) offer extremely stable and precise measurements due to their inherently large scale factor and fixed atomic frequency reference. This makes them excellent candidates for inertial navigation, either as part of an IMU or as an external sensor to improve the performance of classical accelerometers via data fusion techniques. State-of-the-art gravimeters have demonstrated sensitivities of $2 \times 10^{-10}\,g$ after 1000 s of integration using interrogation times of $T = 80$ ms \cite{Gillot-Metrologia-2014,Freier-arXiv-2015}. However, it is challenging to design a compact system with a high sensitivity because the atoms fall with gravity (12.5 cm for $2T = 160$ ms). Nevertheless, intermediate performances at smaller length scales can still be of interest since the position accuracy $\delta r$ of classical sensors is largely determined by the amplitude of Schuler oscillations \cite{Schuler-PhysikZeit-1923}
\begin{equation}
  \label{eq:Schuler}
  \delta r \simeq R_{\oplus} \frac{\delta a}{g},
\end{equation}
where $R_{\oplus} = 6371$ km is the mean radius of the Earth, $\delta a$ is the bias stability of the accelerometer, and $g$ is the acceleration due to gravity. For instance, an accelerometer with a short-term sensitivity of $1 \, \mu g$, and a long-term bias stability of $0.1 \, \mu g$ will reduce the amplitude of the Schuler oscillation to 6 m, which is one order of magnitude below standard navigation technology. We believe this is reachable with interrogation times of $T \lesssim 10$ ms, which can be accomplished with a compact device.

CAAs have the potential to surpass both the short-term sensitivity and long-term stability currently available in most classical technologies \cite{Gillot-Metrologia-2014}. The first onboard CAA experiments \cite{Geiger-NatComm-2011} demonstrated a modest short-term (1 s) sensitivity of $2 \times 10^{-4}\,g$ at $T = 1.5$ ms, which corresponds to roughly $3 \times 10^{-6}\,g$ after 1 h by extrapolation. This is already competitive with strategic-grade accelerometers, as indicated in \Tab \ref{tab:SensorGrades}. However, there are still many challenges to overcome before this technology can be transferred to a commercial device.

For inertial navigation, a dynamic range of $2g$ is usually sufficient for most applications. This is one major drawback of cold-atom-based sensors. Since the interferometer output is periodic, the acceleration range is intrinsically limited to $\pi/k_{\rm eff} T^2$ (corresponding to the invertable range of sinusoidal functions). External accelerations of the reference mirror can cause the fringe to shift by more than $\pi$ rad, which can corrupt the measurement output of CAAs. For instance, at $T = 3$ ms a half fringe corresponds to only $\pi/k_{\rm eff} T^2 \sim 2 \times 10^{-3}\,g$. This necessitates measurements from auxilary accelerometers to re-center the interferometer output on the correct fringe \cite{Geiger-NatComm-2011, Barrett-NJP-2015, Lautier-APL-2014}. Hybridization techniques like the one we discuss in \Sec \ref{sec:HorizontalAxis} allow one to extend the dynamic range of CAAs to that of a classical sensor. However, they also impose stringent requirements on the resolution of the mechanical accelerometer \cite{Barrett-NJP-2015}.

The interest of cold-atom gyroscopes (CAGs) can be understood with an analogy of the phase shift induced by the Sagnac effect
\begin{equation}
  \label{eq.Sagnac}
  \Delta\Phi_{\rm Sagnac} = \frac{4\pi E}{h c^2} \bm{A} \cdot \bm{\Omega},
\end{equation}
where $\bm{A}$ is the area vector of the Sagnac loop and $E$ the energy of the particle ($E = hc/\lambda$ for a photon of wavelength $\lambda$, and $E = M c^2$ for a particle with rest mass $M$). As a consequence of the larger internal energy of an atom, the scale factor for the Sagnac phase is $\sim 10^{11}$ higher for a matter-wave interferometer with the same area as an optical one. A characteristic rotation rate is $\Omega_{\pi}$, \ie the rate rotation that produces a phase shift of $\pi$. Thus, a CAG requires a $10^{11}$ smaller enclosed area to measure $\Omega_{\pi}$ compared to an optical gyroscope---meaning one can envision a very compact system that can measure small rotation rates. It is even possible to increase the area of the atomic gyroscopes by utilizing large momentum transfer pulses to separate the arms of the interferometer \cite{Muller-PRL-2008, Chiow-PRL-2011}. For instance, if the atomic beam-splitters are composed of high-order Bragg pulses which transfer $N \hbar k_{\rm eff}$ to the atoms, the area of the CAG increases by a factor $N$ without significantly increasing size of the apparatus. An equivalent technique is used with optical gyroscopes where the light can be recirculated many times, for example in an optical fiber loop, yielding an $N$-fold increase in the enclosed area.

An entirely equivalent picture considers a CAG as an accelerometer that measure the Coriolis acceleration. However, compared with a CAA, the atoms must be launched with a velocity $\bm{v}$ transverse to the laser wavefronts. Two general categories of CAGs have been demonstrated so far: those using transversely-cooled atomic beams \cite{Gustavson-PRL-1997, Gustavson-ClassQuantumGrav-2000, Durfee-PRL-2006}, and those using cold atoms launched in a moving molasses \cite{Canuel-PRL-2006, Gauguet-PRA-2009, Tackmann-NJP-2012, Tackmann-CRPhys-2014, Berg-PRL-2015}. Both categories have shown competitive sensitivities. Similar to CAAs, they have demonstrated impressive performance levels of $\sim 10^{-7}$ rad/s at short term (1 s) and a few $10^{-9}$ rad/s at long term (1 h) \cite{Barrett-CRPhys-2014}, but require very large vacuum systems to launch the atoms. The ultimate goal of these devices would be to reach long term bias stabilities of $10^{-6}\,\Omega_{\oplus} \sim 10^{-10}$ rad/s, where $\Omega_{\oplus} \simeq 7.3 \times 10^{-5}$ rad/s is the rotation rate of the Earth. This corresponds to an order of magnitude compared to the state-of-the-art of both FOGs and CAGs. The required dynamic range depends on the application, but can vary from $4 \Omega_{\oplus}$ to $10^5 \, \Omega_{\oplus}$ \cite{Lefevre-Book-2014}. The typical dynamic range for a CAG is equal to $\Omega_{\pi} \sim \Omega_{\oplus}/3$. One can imagine applying the same hybridization techniques with a CAG as for a CAA, for example, by replacing the classical accelerometer with a FOG.

One issue that CAIs face is a loss of information from the noise spectrum due to the dead time between measurements that is required to prepare each cold sample. This loss can directly affect the sensitivity via an aliasing effect. Nevertheless, the loss of sensitivity can be ameliorated by using joint operation techniques \cite{Meunier-PRA-2014}, or by using high-repetition rates ($\sim 100$ Hz) with less dead time by recapturing the atoms \cite{McGuinness-APL-2012, Rakholia-PRA-2014}. High dynamic movements and outage issues (\ie if the motion is too extreme for the sensor) have been studied theoretically in the context of a hybrid cold-atom-based IMU and classical navigation-grade IMU \cite{Canciani-Thesis-2012, Jekeli-JInstNav-2005}. Even if the signal from the atomic IMU is not always available, it is still possible to improve the performance of a classical IMU using data fusion techniques.

\section{Onboard cold atom accelerometer}
\label{sec:OnboardICE}

The core challenge faced when operating a CAA onboard a moving vehicle is that, due to the high sensitivity of the device, even low levels of vibrations can overwhelm the measurements of the acceleration of the atoms. Since the interferometer signal has a non-linear periodic response (which limits its dynamic range), the interference fringes are ``blurred'' by the random accelerations caused by vibrations. To overcome this problem, it is necessary to extend the dynamic range of the atomic sensor---making it possible to deduce the acceleration over many fringes \cite{Geiger-NatComm-2011, Barrett-NJP-2015}, or to dynamically lock the fringe on a strongly varying acceleration signal \cite{Lautier-APL-2014}.

\subsection{Measurement of the acceleration along a horizontal axis}
\label{sec:HorizontalAxis}

The ICE experiment is one of the first projects to have measured horizontal accelerations for inertial navigation applications \cite{Stern-EPJD-2009, Geiger-NatComm-2011}. Most other atom interferometer experiments are designed to measure the acceleration along the vertical axis, such as gravimeters where the goal is to make precise measurements of $g$ for various applications. In these cases, mirror vibrations are highly damped by passive or active isolation stages. On the contrary, for inertial navigation, the small accelerations caused by vibrations can be an important part of the signal needed to determine the position of an object, and therefore should not be rejected. They also constitute a strongly varying signal, and make it challenging for CAAs to operate in a precise manner. Additional difficulties arise in the specific regime of low velocity and low acceleration, for which the usual techniques used in atomic gravimeters cannot be applied (\ie the counter-propagating transition frequencies are degenerate for both $\pm \bm{k}_{\rm eff}$ directions).

\begin{figure}[!tb]
  \centering
  \includegraphics[width=0.5\textwidth]{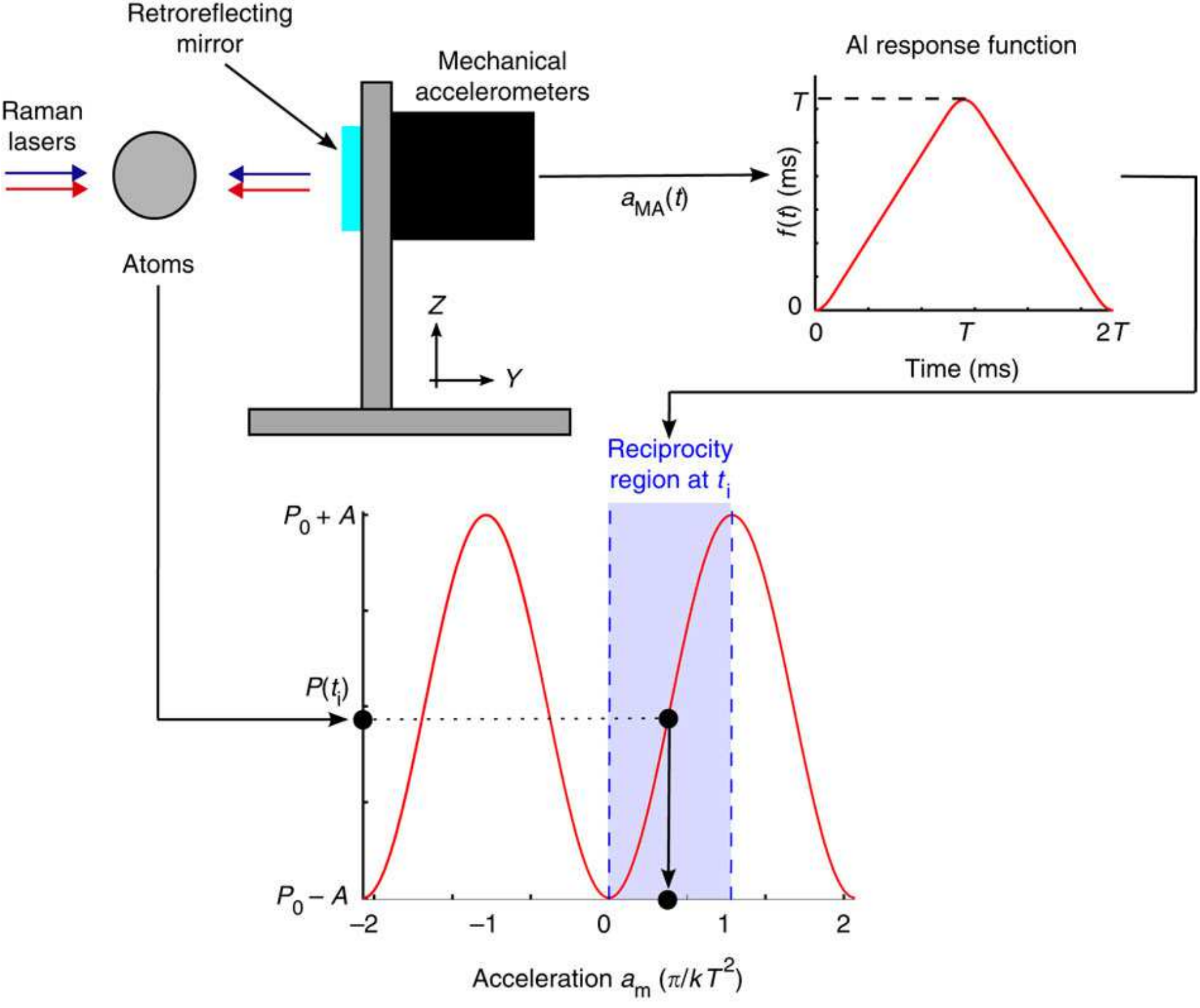}
  \caption{The signal of the mechanical accelerometer is filtered by the response function $f(t)$ of the CAA \cite{Geiger-NatComm-2011, Barrett-NJP-2015}, and gives the reciprocity region where the interferometer operates at the measurement time $t_i = i T_{\rm cyc}$, where $T_{\rm cyc}$ is the cycle time of the experiment. In this example, the reciprocity region corresponds to the $0 - \pi/k_{\rm eff} T^2$ interval. The value provided by the CAA, $P(t_i)$, is then used to refine the acceleration measurement within the reciprocity region. The acceleration is obtained by inverting the response $P(t_i)$ within this region (red curve).}
  \label{fig.HybridICE}
\end{figure}

\begin{figure}[!tb]
  \centering
  \includegraphics[width=0.95\textwidth]{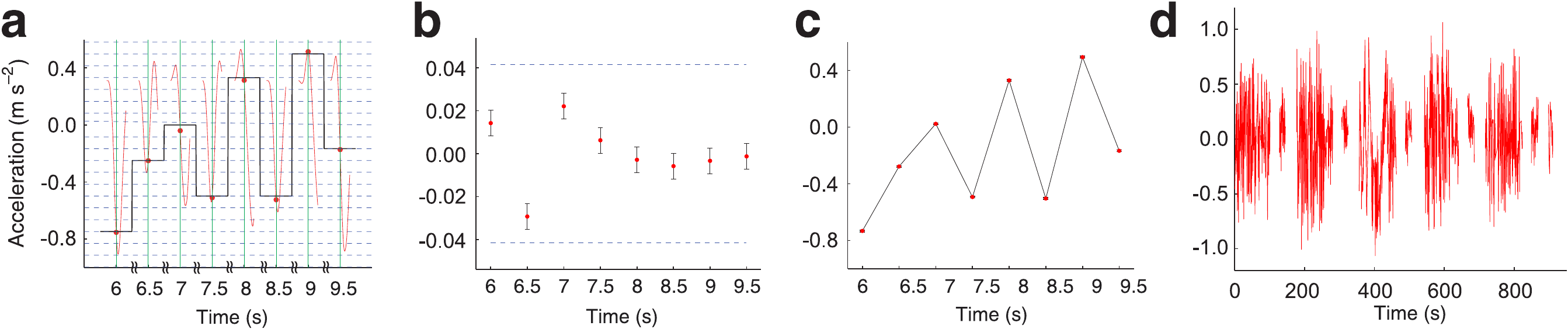}
  \caption{(a) In red, the signal recorded by the mechanical accelerometer in the time window $\{t_i - 2T, t_i + 2T\}$ around the measurement times $t_i = i T_{\rm cyc}$ (vertical green lines), with $T_{\rm cyc} = 500$ ms and $T = 1.5$ ms. The accelerometer signal has been filtered by the response function $f(t)$ of the interferometer \cite{Geiger-NatComm-2011, Barrett-NJP-2015}, and the red points represent the value of the signal at $t_i$. This value determines the reciprocity region where the interferometer operates, delimited by two horizontal dashed lines. In this way, the accelerometer provides a coarse measurement (black step-like signal). (b) The CAA is then used for the high-resolution measurement within its reciprocity region, bounded by the two blue dashed lines at $\pm \pi/2k_{\rm eff} T^2 \simeq \pm 0.087$ m/s$^2$. The error bars represent the noise of the CAA, which is 0.0065 m/s$^2$ per shot in this example (SNR = 4.3). (c) The total acceleration is the sum of the black step-like signal in (a) and the CAA measurements in (b). (d) Full signal measured by the hybrid sensor onboard the Novespace A300 Zero-G aircraft during successive $1g$ and $0g$ phases of flight. For this data set, the resolution of the sensor in 1 s is more than 100 times below the acceleration fluctuations of the aircraft.}
  \label{fig.MeasureVibrationAvion}
\end{figure}

During parabolic flight campaigns in 2010, we measured the horizontal acceleration of the aircraft with a mobile CAA utilizing a laser-cooled sources of $^{87}$Rb atoms \cite{Geiger-NatComm-2011}. Our best performance in this configuration was obtained with an interrogation time of $T = 3$ ms, a signal-to-noise ratio of $\rm{SNR} = 3.1$ and a cycling time of $T_{\rm cyc} = 500$ ms, which gives a one-shot sensitivity of $\delta a = 1/\mbox{SNR} \, k_{\rm eff} T^2 = 2.3 \times 10^{-4}\,g$. To achieve this, we recorded the acceleration of the reference mirror using a mechanical accelerometer, and convolved it with the response function of the CAA to obtain an estimate of the inertial phase, $\Delta\Phi$. The accuracy of this estimate was sufficient to locate the half-fringe number. Then we inverted the output of the CAA, $P(t_i)$, within this half fringe to obtain a more precise value of the acceleration, as shown in \Fig \ref{fig.HybridICE}. Using the same principle as a Vernier scale, we summed the coarse output of the mechanical accelerometer with the precise value estimated from the CAA. This process is illustrated in \Fig \ref{fig.MeasureVibrationAvion}. In this way, we were able to measure the horizontal acceleration of the plane during a full flight, with alternating $1g$ and $0g$ maneuvers, as shown in \Fig \ref{fig.MeasureVibrationAvion}(d). The discontinuity in the measurement is due to the $2g$ phase on either side of the $0g$ phases, where the CAA is not designed to operate. More details can be found in \Ref \citenum{Geiger-NatComm-2011}.

In principle, this technique can be extended to higher precision with larger $T$ until the point where the resolution of the mechanical accelerometer limits the determination of the interferometer half-fringe number. This limit occurs when the phase shift induced by the self-noise of mechanical accelerometer, $S_{\rm self}$ given in $g/\sqrt{\rm Hz}$, reaches $\sim \pi/2$. A more precise calculation based on the frequency response of the interferometer \cite{Cheinet-IEEE-2008, Barrett-NJP-2015} yields $\phi_{\rm self} = \tfrac{1}{\sqrt{3}} k_{\rm eff} S_{\rm self} T^{3/2} < \pi/2$. State-of-the-art mechanical accelerometers exhibit self-noise levels of $S_{\rm self} < 10^{-7}\,g/\sqrt{\rm Hz}$, meaning $T$ could potentially be extended beyond 300 ms before reaching this limit. However, onboard the aircraft we found that the interrogation time was limited by a loss of contrast due to vibrations. This is caused by the random accelerations of the reference mirror which induce a significant shift in the laser frequency seen by the atoms $\delta_{\rm vib}$ due to the Doppler effect
\begin{equation}
  \label{eq.VibDoppler}
  \delta_{\rm vib}(t) = \bm{k}_{\rm eff} \cdot \int_{t_i}^t \bm{a}_{\rm vib}(t') \dd t'.
\end{equation}
Obviously, this time-dependent shift depends on the nature of the vibration spectrum and the particular time interval chosen. We characterize this effect by the maximum shift over the interval $\{t_i, t_i + 2T\}$, given by $\delta^{\rm max}_{\rm vib} = 2 k_{\rm eff} a^{\rm max}_{\rm vib} T$, where $a^{\rm max}_{\rm vib} \simeq 0.5$ m/s$^2$ is the maximum vibration amplitude measured on the Zero-G plane. Thus, at $T = 10$ ms we find $\delta^{\rm max}_{\rm vib} \simeq 160$ kHz, which is large when compared with the Rabi frequency $\Omega_{\rm eff} \simeq 25$ kHz, which determines the ``bandwidth'' of the Raman pulse. If the laser is off-resonant during one of the Raman pulses, the atoms are not efficiently excited and consequently the interferometer contrast decreases. A quantitative model for this loss of contrast will be published elsewhere \cite{Barrett-2016}.

\subsection{Measurement of the acceleration along the vertical axis}
\label{sec:VerticalAxis}

For inertial navigation, one is required to measure the acceleration along three orthogonal directions. After our initial experiments along the horizontal pitch axis of the Zero-G aircraft (which were not sensitive to $g$), we recently carried out the first CAA measurements along the vertical axis of the plane \cite{Barrett-2016}. This required some subtle changes to be made to the experiment. First, during steady flight at $1g$, gravity produces a significative acceleration of the atoms along the measurement axis which shifts the resonance frequency. To account for this we used the same technique that is used in atomic gravimeters---we applied a linear chirp to the Raman laser frequency to stay on resonance with the atoms. However, during the microgravity environment produced by parabolic flight, the mean Doppler shift is negligible compared with $1g$---hence the frequency chirp was disabled during these phases of flight and the sensor acted in the exact same fashion as described in \Sec \ref{sec:HorizontalAxis}. This was done automatically by a computer by discriminating the ``state'' of acceleration from the mechanical accelerometer attached to the reference mirror. During the flight, the CAA runs autonomously, where the user is required only to change parameters based on real-time feedback from the atomic sensor.

\begin{figure}[!tb]
  \centering
  \includegraphics[width=0.7\textwidth]{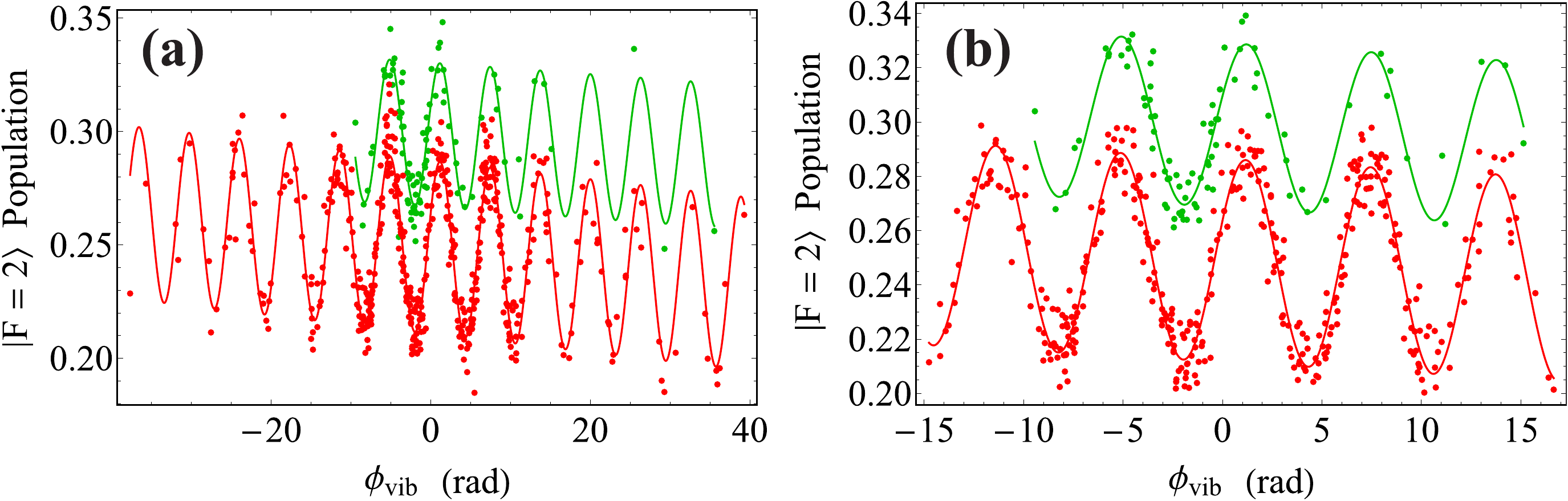}
  \caption{Interference fringes from the mobile $^{87}$Rb interferometer at $T = 1$ ms measured along the vertical axis of the Zero-G aircraft. Here, $\phi_{\rm vib}$ represents the vibration phase computed from measurements given by the mechanical accelerometer, and the vertical axis is the normalized atomic population in the $F = 2$ ground state of rubidium. (a) Raw data in $1g$ (red, SNR $= 6.7$) and $0g$ (green, SNR $= 7.1$). (b) Same as (a) with the data filter applied ($1g$: SNR $= 8.0$, $0g$: SNR $= 7.5$). Data were acquired during five successive parabolas.}
  \label{fig.FringesVertical}
\end{figure}

The central idea behind these experiments is the same as described in \Sec \ref{sec:HorizontalAxis}. By using the fringe reconstruction by accelerometer correlation (FRAC) method described in \Refs \citenum{Geiger-NatComm-2011} and \citenum{Barrett-NJP-2015}, we estimate the vibration-induced phase $\phi_{\rm vib}$ as shown in \Fig \ref{fig.FringesVertical}. In these data, the amplitude noise on the fringes can be explained by a variation of contrast during the flight, which is caused by drifts in laser power between successive parabolas. The linear offset on the fringes is related to movement of the atom cloud during the interferometer, which affects the collection efficiency of fluorescence photons on our 1 mm$^2$ detector. We emphasize that it is unusual for CAIs to scan so many fringes because of vibrations. Typically, atom interferometers are operated in quiet and controlled environments, where vibrations are highly damped.

We have developed a method to improve the signal-to-noise ratio (SNR) of the interference fringes measured under ``noisy'' conditions. As mentioned in \Sec \ref{sec:HorizontalAxis}, residual mirror motion can Doppler-shift the Raman laser off of resonance. However, using measurements from the mechanical accelerometer, it is possible to test if a given repetition of the experiment satisfies the resonance condition or not. Using the results of these tests, we filter the data in post treatment. A ``good'' data point must pass three criteria, one for each Raman pulse. For the first $\pi/2$-pulse, the Doppler effect must be smaller than the characteristic width of the velocity distribution ($\sigma_v$) such that the selected velocity class defined by the Raman frequency contains a sufficient number of atoms. This criteria can be written as
\begin{equation}
  \label{eq.DopplerInsideVelocitydistrib}
  |\delta_{\rm las} - \delta_{\rm vib}(t_1)| < k_{\rm eff} \sigma_v,
\end{equation}
where $\delta_{\rm las}$ is the Raman laser detuning compared to the two-photon transition, and $\delta_{\rm vib}(t_1)$ is the vibration-induced Doppler shift defined by \Eq \eqref{eq.VibDoppler} at the time of the first pulse, $t_1$. The second and third criteria determine whether or not the Doppler shift is sufficiently inside the Raman pulse bandwidth ($\Omega_{\rm eff}$) at the time of the second and third Raman pulses, respectively:
\begin{equation}
  \label{eq.DopplerInsideOmegaRabi}
  |\delta_{\rm las} - \delta_{\rm vib}(t_i)| < \frac{\Omega_{\rm eff}}{4},
\end{equation}
where $t_i$ is the instant of the $i^{\rm th}$ pulse. This filter is effective at removing ``bad'' points from the data, as shown in \Fig \ref{fig.FringesVertical}(b), which improves the SNR of the measured fringes.

\begin{figure}[!tb]
  \centering
  \includegraphics[width=0.95\textwidth]{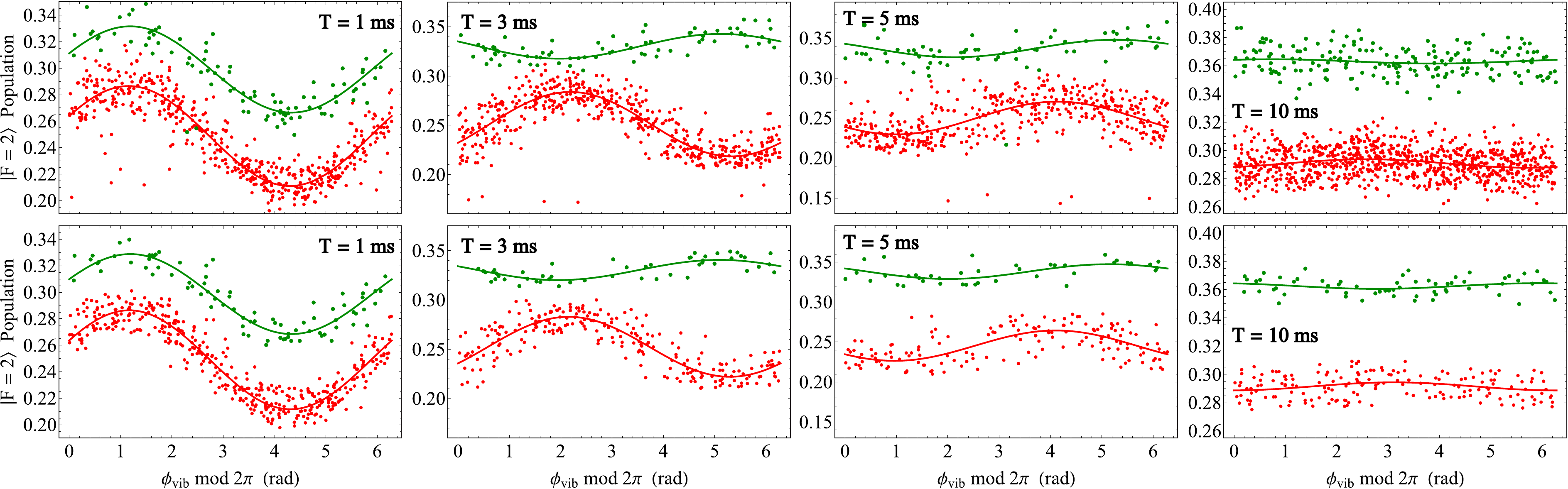}
  \caption{Fringes measured along the vertical direction in the Zero-G aircraft for various interrogation times ($T = 1$, 3, 5, and 10 ms). Here, plots in the top row indicate raw data and the bottom row show the same data with the vibration filter applied. As in \Fig \ref{fig.FringesVertical}, red points indicate $1g$ and green points indicate $0g$. For visibility all data are folded into the phase range of 0 to $2\pi$ via the modulus function. For clarity, the fringes in $1g$ and $0g$ have been vertically shifted relative to one another.}
  \label{fig.FringesVerticalFold}
\end{figure}

Figure \ref{fig.FringesVerticalFold} displays data acquired under similar conditions to those presented in \Fig \ref{fig.FringesVertical}(a) for various interrogation times. Here, we have plotted the vibration phase modulo $2\pi$ so that the phase span is folded into the range of $0 - 2\pi$ for all values of $T$. In fact, the number of fringes spanned by the vibrations increases in proportion to $T^2$. For instance, with $T = 10$ ms and $a_{\rm vib} = \pm 0.5$ m/s$^{2}$, the range of phase is $\pm 800$ rad. Here, it is remarkable that fringes are still visible, which is a measure of the quality of correlation between the CAA and the mechanical accelerometer over such a large dynamic range.

When oriented along the vertical direction of the aircraft the interferometer is sensitive to rotations about the horizontal axes. These rotations can cause a phase shift due to the Coriolis acceleration in the rotating frame, and also a loss of fringe contrast because the two interferometer pathways do not perfectly close \cite{Barrett-2016, Lan-PRL-2012}. Both of these effects are larger during a parabolic maneuver than in steady flight since the rotation rate is substantial (4 deg/s $\sim 0.07$ rad/s). From \Fig \ref{fig.FringesVerticalFold}, it is clear that the loss of contrast increases with $T$ for both data in $1g$ and $0g$. However, in $1g$ this loss is dominated by the Doppler shift due to vibrations, while in $0g$ we estimate that the contrast loss is dominated by rotations. Obviously both effects severely limit the timescale and sensitivity of the interferometer, and future generations of mobile CAAs will need to address this problem.

\subsection{Three-axis inertial sensor}
\label{sec:3DCAI}

As illustrated by the ICE experiment, hybridization between classical and quantum sensors allows us to benefit from the advantages of both technologies (\ie high dynamic range and continuous measurement output from the classical device, and high sensitivity and long-term stability from the atomic sensor). So far, we have combined a three-axis mechanical accelerometer with a single-axis atomic accelerometer. In the future, this hybridization scheme will be extended to include fiber-optic gyroscopes to measure the three axes of rotation and acceleration. This will allow us to fully reconstruct the phase of the single-axis interferometer, and eventually envision a complete cold-atom-based IMU.

Historically, atom interferometers have been designed to be sensitive to inertial effects along only one axis---namely the one defined by the interferometry laser beams. This has been motivated primarily by convenience and the specific measurement application of the interferometer. However, there is no fundamental limitation preventing one from expanding to two or three dimensions \cite{Canuel-PRL-2006, Antoine-PhysLettA-2003, Borde-CRAcadSciParis-2001}. Unlike the internal spin states of the atom, which are sensitive to a particular quantization axis, external momentum states can be used simultaneously along any direction to coherently store phase information \cite{Freegarde-PRL-2003}. An atom interferometer functions by coherently splitting and recombining matter waves that have traveled along two spatially separated pathways. The idea is to be able to separate the scalar phase $\Delta\Phi$ given by \Eq \eqref{eq.PhiAccelero} into three independent components that compose a \emph{vector} phase $\Delta\bm{\Phi} = \{\Delta\Phi_x, \Delta\Phi_y, \Delta\Phi_z\}$---one along each direction:
\begin{equation}
  \label{eq.phaseXYZ}
  \Delta\Phi_{x,y,z}
  = |\bm{k}_{\rm eff}| \bm{\epsilon}_{x,y,z} \cdot (\bm{r}_1 - 2\bm{r}_2 + \bm{r}_3) + \Delta \phi
  = k_{\rm eff} a_{x,y,z} T^2 + \Delta \phi,
\end{equation}
where $\bm{\epsilon}_{x,y,z}$ is the unit vector along each orthogonal direction and the vector $\bm{r}_i$ represent the center-of-mass positions of the atomic wavepacket during each of the three laser pulses. The vector phase $\Delta\bm{\Phi}$ is then proportional to the vector acceleration $\bm{a}$ of the atoms in 3D.

\begin{figure}[!tb]
  \centering
  \includegraphics[width=0.6\textwidth]{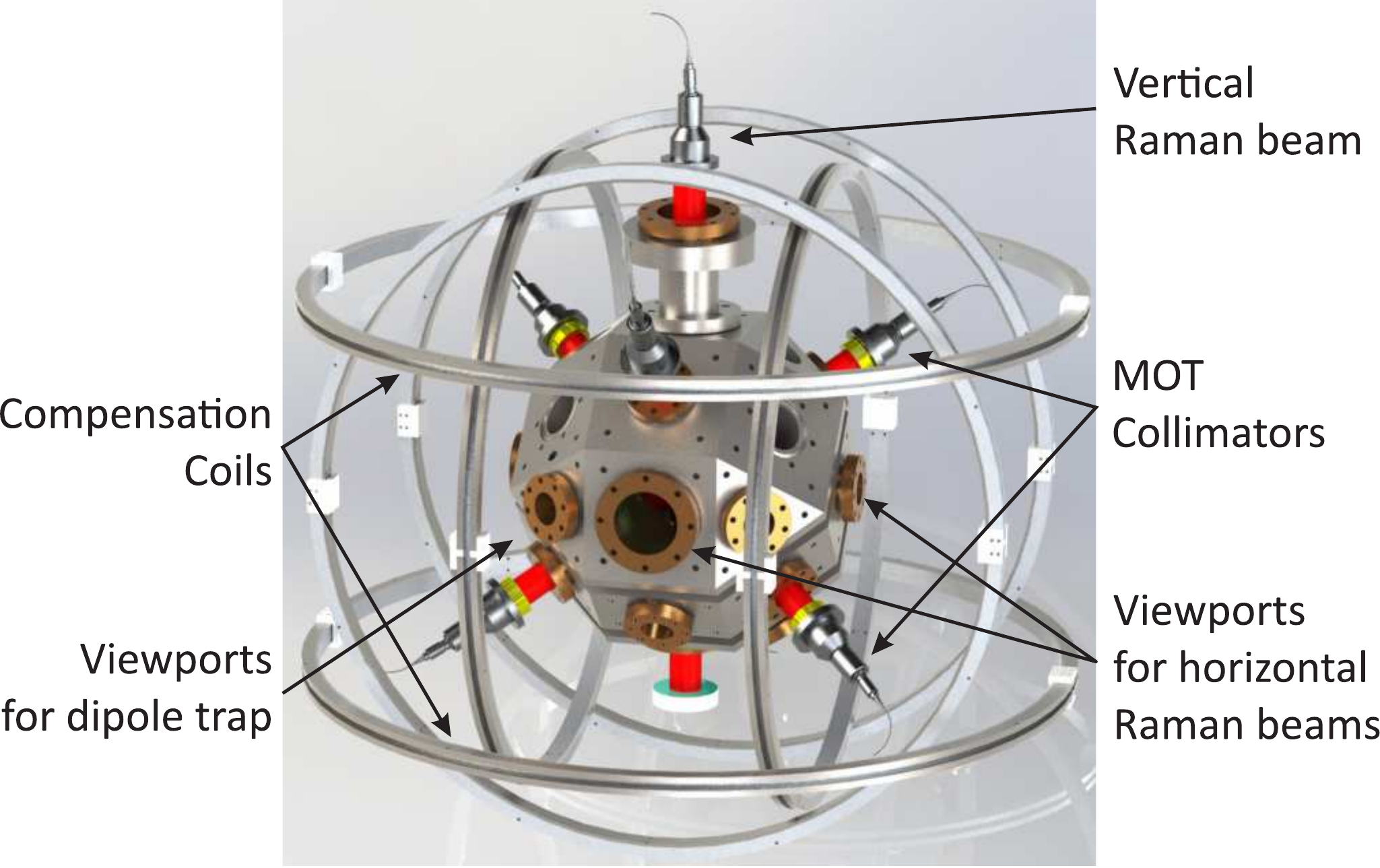}
  \caption{Apparatus built on the ICE experiment to achieve a 3D CAA.}
  \label{fig.ICEVacuum}
\end{figure}

The ICE experiment is designed to achieve this three-axis measurement. After laser-cooling the atoms in a vapor-loaded magneto-optical trap, they will be loaded into an optical dipole trap and evaporated to ultra-cold temperatures. To implement the three-axis sensor, three independent pairs of Raman beams will be aligned along the $x$, $y$ and $z$ axes, through the center of the atoms, and are retro-reflected off of separate mirrors (see \Fig \ref{fig.ICEVacuum}). The ultra-cold source will be spatially localized at the cross section of the three Raman beams. These beams can be pulsed on either simultaneously or sequentially in a $\pi/2 - \pi - \pi/2$ sequence to split, reflect and recombine the atoms. This will enable measurements of the vector acceleration $\bm{a}$. Detection of the interference will combine two CCD cameras and two fluorescence imaging beams, imaging two orthogonal planes to acquire sufficient information to reconstruct the interference \cite{Sugarbaker-PRL-2013}. This will allow us to spatially-discriminate between inertial phase shifts along the $x$, $y$ and $z$ axes. A mechanical accelerometer will be rigidly fixed to the back of each retro-reflection mirror to measure vibrations. These vibration measurements will be converted to phase shifts of the interferometer along each axis, and correlated with the atomic populations measured by fluorescence imaging in the same spirit as described in the previous sections.

\section{Fast tunable laser source for compact cold-atom sensors}
\label{sec:FastTunableLaser}

One of the most important and complex parts of cold-atom-based systems is the laser source. Typically, these systems demand control of the optical frequency with an accuracy of the order of 100 kHz, and a tuning range of $\sim 1$ GHz around the atomic transition. Fast tunability over this range in less than 1 ms is also desired to independently optimize the different phases of a measurement sequence (\eg cooling, preparation, interrogation and detection). In this section we describe a compact laser system for cooling and manipulating $^{87}$Rb. The architecture is based on phase modulating a single laser diode operating in the telecom frequency band. With this design, the laser frequency can be tuned over $\sim 1$ GHz on timescales of a few hundred $\mu$s by precisely controlling the frequency sidebands created by a modulation signal. While the laser source presented here has been specifically developed for the MiniAtom project \cite{MiniAtom-Website}, the concepts can be used for many different cold-atom experiments. To improve the robustness and reliability of the source, the laser is composed primarily of telecom fibered components. Indeed, standard laser systems in laboratories based on semiconductor diodes in the optical domain require many free-space components and are consequently bulky and sensitive to vibrations and temperature fluctuations. Moreover, the lifetime of semiconductor lasers and amplifiers is not better than a few years. In comparison, telecom components at 1560 nm are natural candidates for compact and transportable devices since they are all fibered, compact and very robust. They satisfy Telcordia standards---meaning they have been certified for demanding environments and have a lifetime longer than 25 years. Additionally, maintenance and upgrade are easy, and market availability for the next 50 years is assured. This technology has already been tested on the ICE experiment in the A300 Zero-G airbus at vibration levels of up to 0.5 m/s$^2$ rms \cite{Menoret-OptLett-2011}. Moreover, telecom components are routinely used in commercial FOGs \cite{Lefevre-Book-2014} and are promising for space applications \cite{Schuldt-ExpAst-2015}.

The laser system presented in this section, including electronics and power supplies, has been developed for the architecture of an atomic gravimeter based on a pyramidal reflector, which utilizes a single beam for cooling, preparation, inducing Raman transitions and detection \cite{Bodart-ApplPhysLett-2010}. For this specific architecture only one optical fiber output is required. One challenging aspect of this configuration is that the frequency of this single beam must be changed dynamically during the measurement sequence, which relies critically on the fast tunability of the laser source.

\subsection{Architecture}
\label{sec:ArchitectureLaser}

\begin{figure}[!tb]
  \centering
  \includegraphics[width=0.5\textwidth]{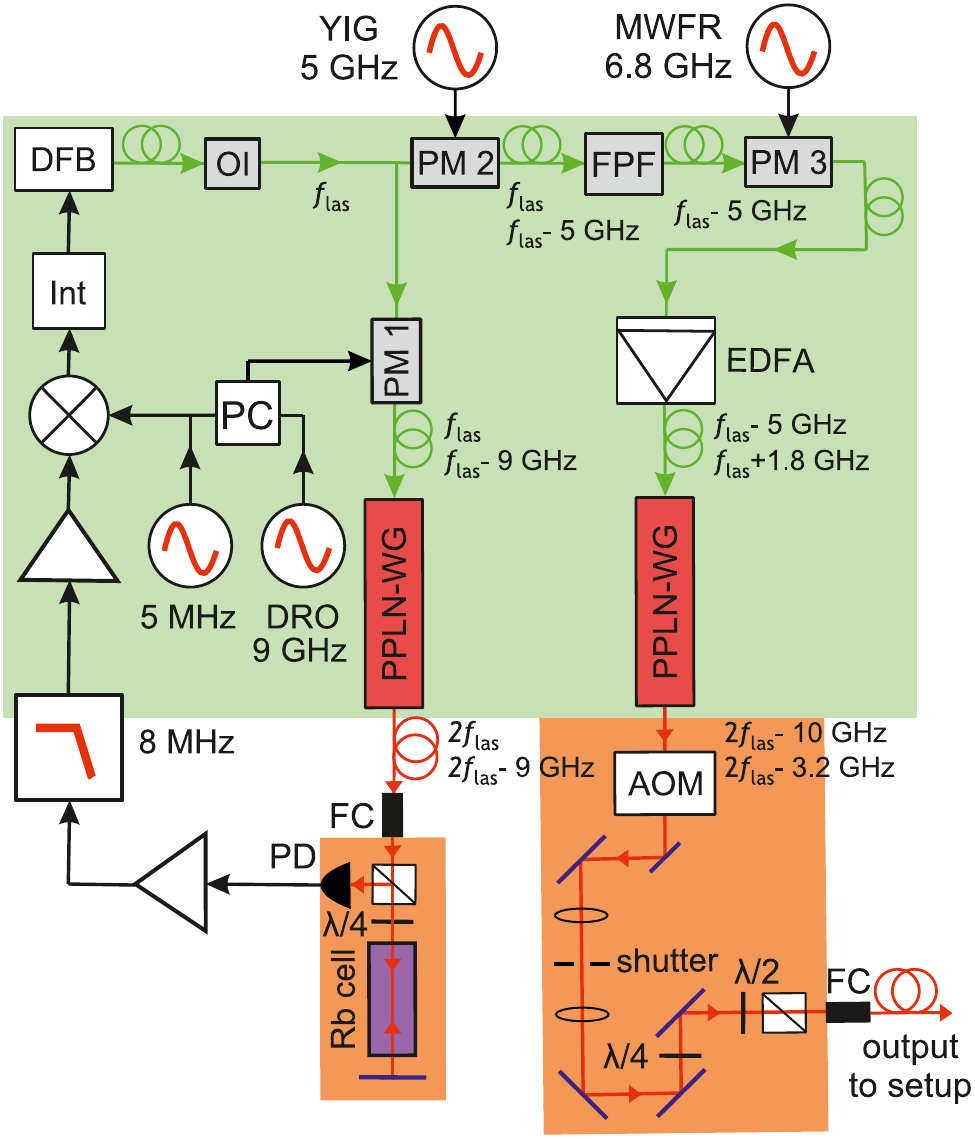}
  \caption{Architecture of the laser source. We use telecom components (green area) to develop a compact fiber-based laser system. A small free-space optical bench (orange area) was implemented with 780 nm optical components to control the frequency of the laser ($f_{\rm las}$) via saturated absorption spectroscopy, to control the power of the laser output via an AOM, and to filter the polarization. OI: optical isolator; PM: phase modulator; FPF: Fabry-Perot filter; EDFA: erbium-doped fiber amplifier; PPLN-WG: periodically-poled lithium-niobate waveguide; MWFR: microwave frequency reference; YIG: yttrium-iron garnet oscillator; DRO: dielectric resonator oscillator; AOM: acousto-optic modulator; Int: integrator; PC: power combiner; FC: fiber coupler; PD: photodiode.}
  \label{fig.architecture}
\end{figure}

The architecture of the laser system is depicted on Figure \ref{fig.architecture}. It is based on telecom technology combined with second harmonic generation (SHG) \cite{Thompson-OptExp-2003} and phase modulation via the electro-optic effect \cite{Carraz-PRA-2012, Theron-APB-2014}. Our approach, which has evolved from similar techniques used in the microwave frequency domain \cite{PopielGorski-Book-1975}, consists of using a single diode laser locked on an atomic resonance, generating the desired optical frequencies by creating multiple sidebands with phase modulators, and then filtering out the undesired ones. Unlike most laser systems for cold-atom experiments, which are typically based on a master/slave architecture, our frequency agility is achieved by adjusting the microwave frequency signal sent to the optical phase modulators.

The laser source is a tunable distributed feedback (DFB) laser diode (40 mW, ITU channel 21, Emcore 1772 DWDM) which has a typical linewidth of 500 kHz\footnote{The linewidth of the DFB diode can be a limitation on the phase noise of an atom interferometer \cite{LeGouet-EPJD-2007}. An alternative solution is the Telcordia-qualified external cavity diode from Redfern Integrated Optics, which exhibits a linewidth of $< 10$ kHz.}, and with a frequency controlled via temperature feedback over a range of 300 GHz. After an optical isolator, a small part of the optical power is diverted, frequency-doubled via SHG and then sent to a saturated absorption spectroscopy module. A dielectric resonator oscillator (DRO) generates a microwave signal of 9.18 GHz, which is amplified to 20 dBm, and sent to the phase modulator PM1 to shift the laser frequency at the input of the spectroscopy module \cite{Peng-OptLett-2014}. We also apply a 5 MHz signal on the same phase modulator to generate sidebands on the input light. The measured absorption signal is later demodulated at the same frequency to obtain an error signal. The laser frequency after SHG, $2f_{\rm las}$, is then servo-locked using the sideband at $2f_{\rm las} - 9.18$ GHz, which is set on resonance with the crossover transition $F = 3 \to F'=3(4)$ of $^{85}$Rb---corresponding to a detuning of $\sim 1$ GHz to the blue of the $F = 2 \to F' = 3$ cycling transition of $^{87}$Rb.

\begin{figure}[!tb]
  \centering
  \includegraphics[width=0.6\textwidth]{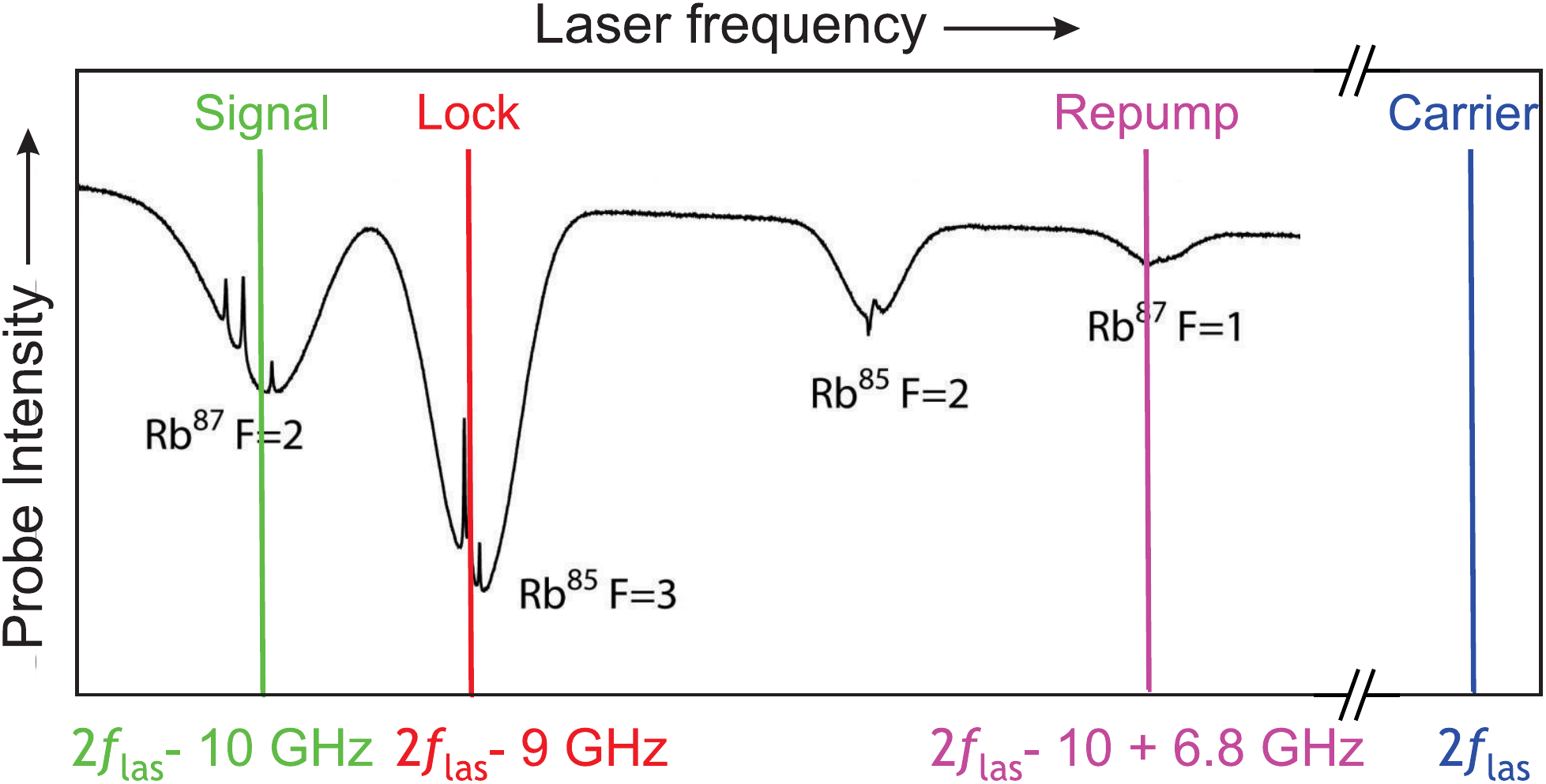}
  \caption{Diagram of the frequencies involved in the architecture of the laser source. A typical saturated absorption spectrum of rubidium is shown for reference. The line labeled ``lock'' corresponds to the $-9$ GHz sideband created by the DRO and PM1, and is locked on the $^{85}$Rb $F = 2 \to 3(4)$ crossover transition. The ``carrier'' line, that is the frequency-doubled output of the laser diode, is off resonance by approximately 3.2 GHz from the nearest transition of $^{87}$Rb. An additional far off-resonant sideband at $2f_{\rm las} + 9$ GHz is also present, but not shown. These frequencies are only present in the saturated absorption part of the setup shown in \Fig \ref{fig.architecture}. The ``signal'' line originates from the $-5$ GHz sideband generated by PM2, which is filtered and frequency-doubled to obtain $2f_{\rm las} - 10$ GHz. This frequency is near the $^{87}$Rb $F = 2 \to 3$ cycling transition. Finally, the ``repump'' frequency at $2f_{\rm las} - 3.2$ GHz is generated by the 6.8 GHz modulation signal sent to PM3, which is near the $^{87}$Rb $F = 1 \to 2$ repump transition. Again, the additional far off-resonant sideband at $2f_{\rm las} - 16.8$ GHz is not shown. Only these frequencies are present on the output fiber shown in \Fig \ref{fig.architecture}.}
  \label{fig.lesraies}
\end{figure}

Figure \ref{fig.lesraies} shows the location of the frequencies generated by each phase modulator. To generate the laser light sent to the atoms, we create a sideband using phase modulator PM2 at $f_{\rm las} - 5.16$ GHz with a YIG oscillator operating at 5.16 GHz. This sideband then transmits through a fibered Fabry-Perot filter (Micron Optics FFP-TF2), and the carrier and other sidebands are suppressed. After SHG, we obtain $2f_{\rm las} - 10.32$ GHz, which yields a detuning of $-2.5\,\Gamma$ relative to the cycling transition of $^{87}$Rb, where $\Gamma \simeq 6$ MHz is the natural linewidth of the atomic transition.

Phase modulator PM3 (10 GHz, 1550 nm, PHOTLINE MPZ-LN-10), located after the Fabry-Perot filter in \Fig \ref{fig.architecture}, creates the repumping frequency that is used for cooling, detection, and as the second Raman frequency for the interferometer. To generate the modulation at 6.8 GHz, a compact microwave source was developed \cite{Lautier-RSI-2014}. This beam is then amplified by an erbium-doped fiber amplifier (EDFA, 27 dBm, Manlight HWT-EDFA-PM-SC-HPC27) yielding 560 mW at 1560 nm. Frequency doubling from 1560 nm to 780 nm is accomplished via SHG in a periodically-poled lithium niobate waveguide (PPLN-WG), where the confinement of the optical mode leads to high intensities and thus high efficiencies\footnote{This is a key element and we found reliable PPLN-WG components from the Japanese manufacturer NTT.}. We typically obtain 280 mW at the output of this module. Since fibered components at 780 nm are not fully reliable from the polarization point of view, to avoid a significant loss of power, we implemented a conservative free-space platform that contains an acousto-optic modulator (AOM) and a mechanical shutter to switch on and off the output light.

\subsection{Description of the prototype}
\label{sec:PrototypeLaser}

The laser source is assembled in a transportable 19-inch 12U rack. The weight of this rack is 80 kg and the consumption of electrical power was evaluated at 200 W. The laser source is powered with a standard AC 220V cord. The source is composed of 4 modules: (i) the frequency control and servo-lock module including the telecom components and the absorption spectroscopy; (ii) the amplification and frequency-doubling stage including the free-space AOM and mechanical shutter, (iii) the AC/DC power converter stage, and (iv) the electronic control stage. The output fiber is a 3 m-long polarization maintaining fiber at 780 nm. The laser source is externally controlled by a computer and a multifunction data acquisition card.

\subsection{Performances of the source}
\label{sec:PerformanceLaser}

\begin{figure}[!tb]
  \centering
  \includegraphics[width=0.5\textwidth]{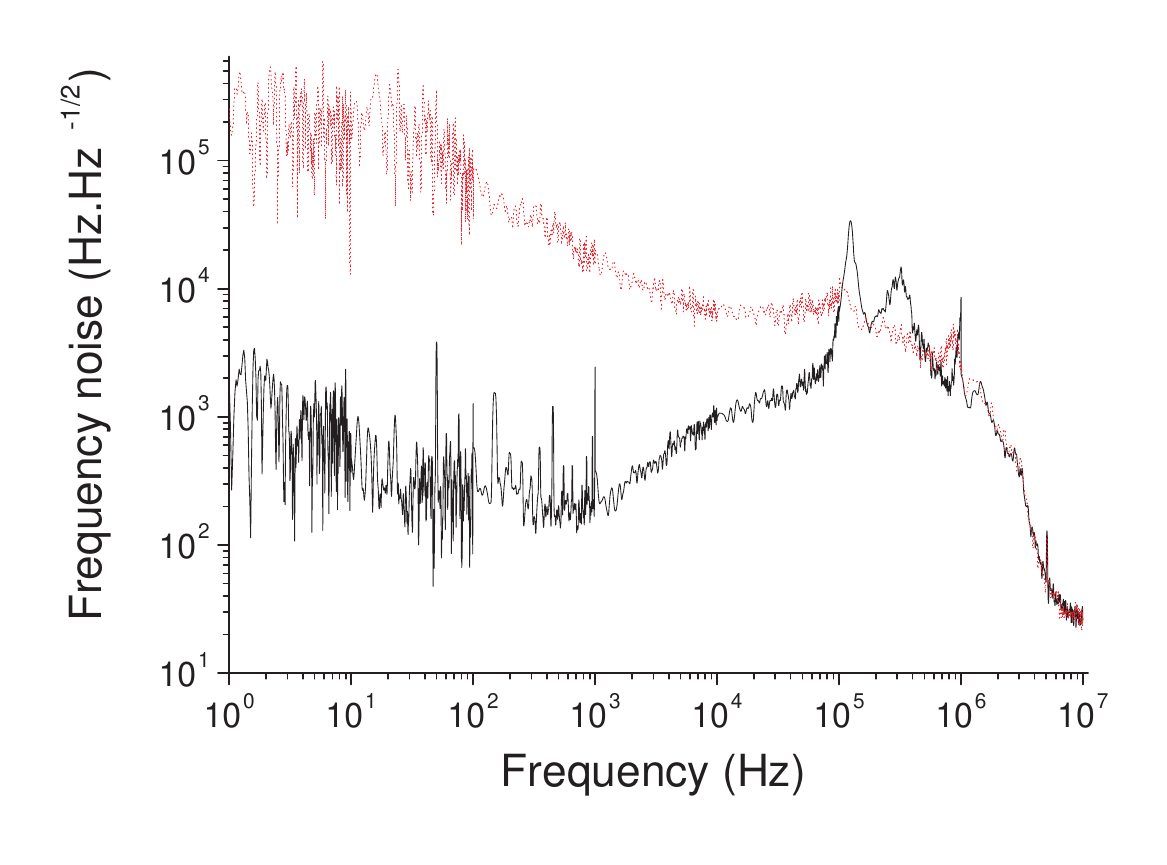}
  \caption{The spectral density of frequency noise of the laser source is measured by a Fourier transform of the error signal in free-running mode (red dots) and with a bandwidth-optimized servo lock (solid black line).}
  \label{fig.locklaser}
\end{figure}

In order to estimate the linewidth of our laser system, we performed a beat-note measurement between our locked laser diode and an integrated external cavity laser diode whose linewidth is specified to be below 15 kHz. The full width at half maximum (FWHM) of the beat note is 500 kHz, which gives us directly the linewidth of the source at 1560 nm. This value is consistent with what we expect for a DFB laser diode and a low-noise current supply. To deduce the linewidth of the source at 780 nm, we take into account the influence of frequency doubling which widens the linewidth\footnote{When doubling the frequency, the linewidth doubles if the laser has a Gaussian spectral lineshape, and quadruples if the laser has a Lorentzian lineshape.}. We measure the spectral density of frequency noise of the laser source in a free-running mode and with a servo lock of optimized bandwidth. Figure \ref{fig.locklaser} shows that the servo lock allows us to reduce the frequency noise of the DFB diode below 100 kHz, which is the relevant band for an atom interferometer \cite{LeGouet-EPJD-2007}. The bandwidth of the lock is currently limited by the diode laser current supply, but in principle it could be increased to $\sim 500$ kHz without significant changes to the architecture.

The tunable range after frequency doubling is $\sim 1$ GHz, and this entire range can be scanned in less then 200 $\mu$s---the speed being limited by the YIG oscillator. In order to achieve the required tuning range, the Fabry-Perot filter should not be too selective (in our case, the finesse is 10 and the free-spectral range is 50 GHz). A shift of the laser frequency can produce a power variation after the filter, thus we can expect a loss of the power of $1\%$ between the center and the edge of the frequency range of the laser, but this effect is negligible at the output since the EDFA is saturated. We chose a configuration which gives priority to the frequency agility, but the drawback is that we suppress sidebands by only a factor 5 and this can have an influence on the sensitivity and accuracy of the atom interferometer \cite{Carraz-PRA-2012}. We can improve this suppression by increasing the finesse of the Fabry-Perot filter and by reducing the tunable range.

The optical power at 780 nm at the output of the fiber is 180 mW. We measure power fluctuations of $10\%$ during the first hour of warm up, then the fluctuations decrease to $1\%$. The polarization is linear along the slow axis of the fiber and the polarization extinction ratio (PER) is 27 dB. We have tested the laser system on the ICE experiment and obtained a magneto-optical trap of $\sim 10^8$ $^{87}$Rb atoms. We also tested the laser on the MiniAtom prototype vacuum system and obtained $2 \times 10^7$ atoms \cite{MiniAtom-Website}.

\subsection{Applications for the compact laser source}
\label{sec:ApplicationLaser}

To conclude, we developed a fast, tunable laser source for a compact cold-atom inertial sensor, which is based on telecom components, fibered frequency doubling, phase modulation and sideband filtering. With our laser architecture, it is possible to scan a 1 GHz range in less than 200 $\mu$s with a sensitivity of 10 kHz, corresponding to the linewidth of the external cavity laser diode whose optical frequency stays fixed. The technology can cover the wavelength range of 1530--1565 nm (765--782.5 nm after SHG). This single-diode architecture benefits from the potential of telecom technology and thus satisfies the criteria of compactness and reliability for field and onboard applications. The principle of generating sidebands and filtering the undesired frequencies, which is commonly used in the microwave domain, has been validated in the optical domain. Moreover, the combination of the frequency shift and the modulation servo lock by using the same phase modulator is a first step toward function hybridization.

We emphasize that telecom fibered components and frequency doubling is suitable for both field sensors and laboratory experiments. Although we have demonstrated the capability of this source for laser-cooling $^{87}$Rb atoms, the principles set out in this manuscript can easily be adapted to many different cold-atom systems using, for instance, $^{85}$Rb or the isotopes of potassium \cite{Stern-AppOpt-2010, Menoret-OptLett-2011}. For example, only two minor modifications are required to operate with $^{85}$Rb: the microwave frequency reference at 6.8 GHz can be replaced with one operating at 3.0 GHz, and the modulation frequency for the sideband lock can be replaced with 10 GHz instead of 9 GHz.

\section{Conclusion}
\label{sec:Conclusion}

We have demonstrated the operation of a compact multi-axis cold-atom accelerometer onboard an aircraft during both steady flight and parabolic maneuvers that produce a weightless environment. We presented measurements of the acceleration along the vertical axis and horizontal axis with one-shot sensitivities of $2.3 \times 10^{-4}\,g$, and we measured a loss of contrast due to the high level of vibrations onboard the aircraft, and the large rotation rates during parabolic flight. In principle, these limitations can be ameliorated by real-time feedback on the Raman laser frequency and dynamically tilting the retro-reflection mirror \cite{Lan-PRL-2012, Hauth-APB-2013, Dickerson-PRL-2013}. In the near future, we hope to achieve simultaneous inertial measurement along two or three directions, which will constitute the next major step toward realizing a cold-atom IMU. 

In parallel, we have developed compact and robust components for these instruments, such as the MiniAtom laser system, which are suitable for many cold-atom experiments or devices. For instance, the sensor head of the ICE experiment contains a 8 L spherical vacuum system (25 cm diameter), and further developments are underway to gain an order of magnitude in compactness. We have also demonstrated a hybridization technique with classical inertial sensors that allowed us to benefit from the advantages of both technologies. Although more involved data fusion methods can be realized, the basic principles used herein illustrate that cold-atom systems can be used to achieve quasi-continuous measurements of accelerations and rotations with high-sensitivity and outstanding long-term stability over a large dynamic range. We anticipate that this technology will play an important role in next-generation inertial navigation systems, tests of fundamental physics, geodesy measurements from Space, or transportable low-cost intermediate-grade gravimeters for ground-based applications.

\acknowledgments

This work is supported by the French spatial agency CNES (Centre National d'Etudes Spatiales) and the European Space Agency. We acknowledge financial support from ANR (grant number: MINIATOM ANR-09-BLANC-0026 and "iQNav: Laboratoire des syst\`{e}mes quantiques pour le positionnement et la navigation inertielle"), CNRS, Triangle de la physique, the D\'el\'egation G\'en\'erale de l'Armement, action sp\'ecifique GRAM (Gravitation, Relativit\'e, Astronomie et M\'etrologie), and the Institut Francilien pour la Recherche sur les Atomes Froids. We would like to thank iXBlue, Kloe, III-V lab and THALES TRT for their involvement in the MiniAtom project. P. Bouyer thanks Conseil R\'egional d'Aquitaine for the Excellence Chair.

\bibliography{report} 
\bibliographystyle{spiebib} 

\end{document}